\documentclass[aps,prc,twocolumn,floatfix,12pts,superscriptaddress]{revtex4}
\usepackage{epsfig}
\usepackage{amsmath}
\usepackage{color}
\usepackage{graphicx}

\begin{document}

\title{Spectra and elliptic flow of thermal photons from full overlap  U+U collisions at RHIC}
\author{Pingal Dasgupta}
\email{pingaldg@vecc.gov.in}
\author{Rupa Chatterjee}
\email{rupa@vecc.gov.in}
\author{Dinesh K. Srivastava}
\email{dinesh@vecc.gov.in}
\affiliation{Variable Energy Cyclotron Centre, HBNI, 1/AF, Bidhan Nagar, Kolkata-700064, India}

\begin{abstract}
We calculate  $p_T$ spectra and elliptic flow  for tip-tip and body-body configurations of full overlap uranium-uranium (U+U) collisions using a hydrodynamic model with smooth initial density distribution and compare the results with those obtained from Au+Au collisions at RHIC. Production of thermal photons is seen to be significantly larger for tip-tip  collisions compared to  body-body collisions of uranium nuclei in the region $p_T >$ 1 GeV. The difference in the results for the two configurations of U+U collisions depends on the initial energy deposition which is yet to be constrained precisely from hadronic measurements. The thermal photon spectrum from body-body collisions is found to be close to the spectrum from  most central Au+Au collisions at RHIC. The elliptic flow parameter calculated for body-body collisions is found to be large and comparable to the $v_2(p_T)$ for mid-central collisions of Au nuclei. On the other hand, as expected, the $v_2(p_T)$ is close to zero for tip-tip collisions. 
 The qualitative nature of the photon spectra and elliptic flow for the  two different orientations of uranium nuclei is found to be independent of the initial parameters of the model calculation. We show that the photon results from fully overlapping U+U collisions are complementary to the results from Au+Au collisions at RHIC.
\end{abstract}

\maketitle

\section{Introduction} 
Anisotropic flow or in particular elliptic flow is one of the key observables used to study the properties of Quark Gluon Plasma (QGP) produced in collisions of heavy nuclei at relativistic energies. Hydrodynamic model with smooth initial density distribution has been used successfully in recent past to study the bulk properties of matter as it simultaneously explains both the spectra and elliptic flow of charged particles~\cite{uli, hydro1}. It has been shown in many interesting recent studies that event-by-event hydrodynamic model with fluctuating initial conditions~\cite{hannu,pt,scott,hannah, sorenson,nex} explains the elliptic flow results even for most central  collisions of heavy nuclei and also the  large triangular flow of hadrons at RHIC and LHC energies ~\cite{alver, flow_phenix,flow_lhc,flow_atlas} both of which were unexplained earlier by  hydrodynamics  with smooth initial density distribution.

Photons are considered as one of the promising probes to study the properties of quark gluon plasma formed in relativistic heavy ion collisions~\cite{phot}. Recent experimental data from 200A GeV Au+Au collisions at RHIC by PHENIX~\cite{phenix_phot} and from 2.76A TeV Pb+Pb collisions at LHC by ALICE~\cite{alice_phot} have reported  excess of direct photon yield over scaled proton-proton collisions. The excess yield in both the cases is attributed to photon radiation from the thermalized QGP and hot hadronic matter.

\begin{figure}
\centerline{\includegraphics*[width=8.0 cm]{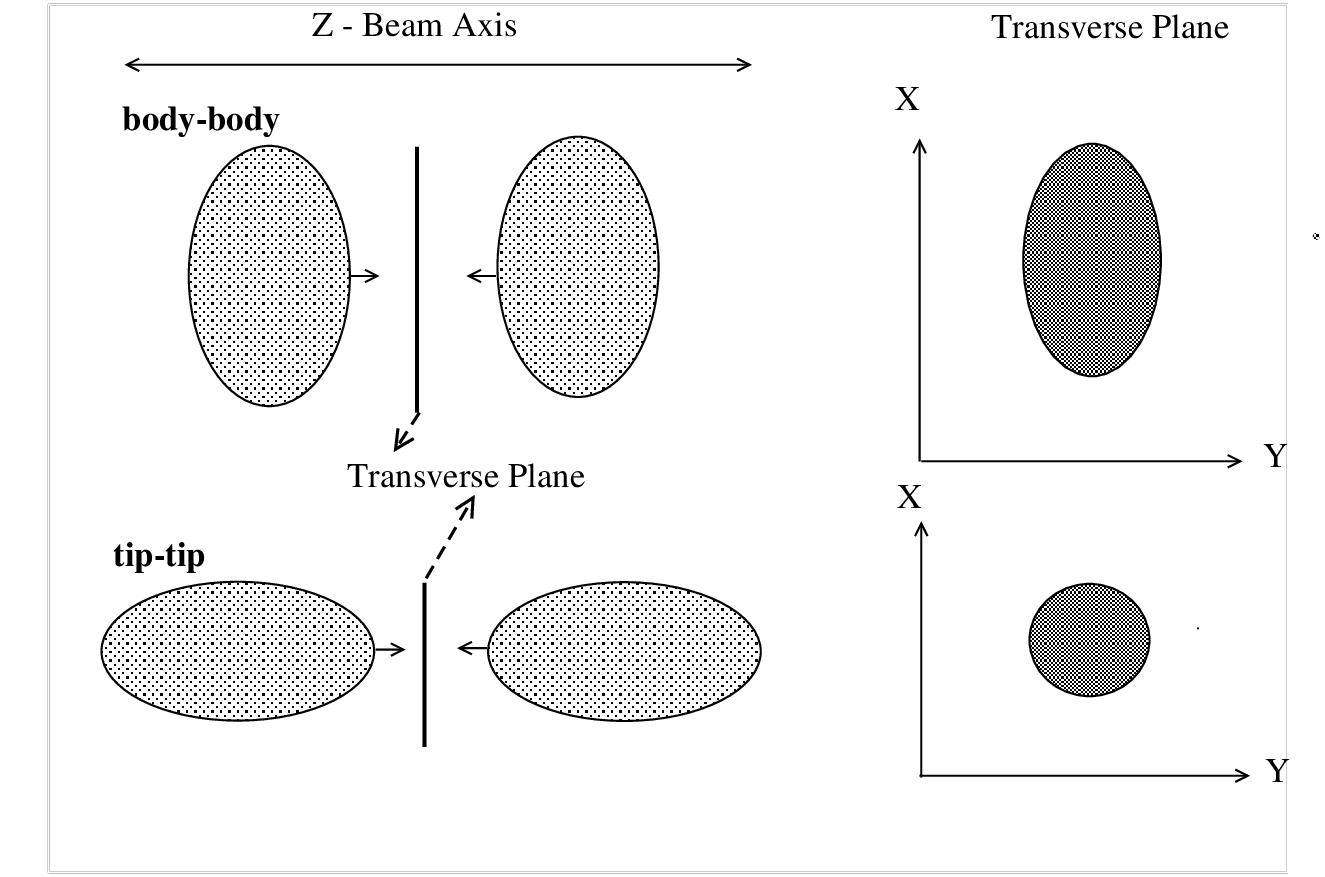}}
\caption{Schematic of tip-tip and body-body collision of full overlap uranium nuclei.}
\label{spec_v2}
\end{figure}

Photon elliptic flow  has the potential to illustrate the hot and dense initial state and its evolution more efficiently compared to hadronic $v_2$~\cite{cfhs}. Direct photon $v_2$ data at RHIC~\cite{phenix_v2} and LHC~\cite{alice_v2} show similar qualitative nature as predicted by model calculations considering hydrodynamical evolution of the system. However, theory results under-estimate the data by a large margin~\cite{chre2}. This is known as the photon $v_2$ puzzle. Many recent studies with  viscous hydrodynamics model using event-by-event fluctuating initial conditions as well as studies considering pre-equilibrium flow have found it difficult to explain the photon spectra and elliptic flow simultaneously. Recent developments in the theory of photon production and calculation of  the photon anisotropic flow parameter in relativistic heavy ion collisions can be found in Refs.~\cite{cs, hannu_phot, phot_v2_hees, maxim, phot_v2_shen, dusling, v22, v23, v24, v25, v26, v27, v28, v29, v30, v31, v32}.

Collisions of uranium  ($^{238}$U) nuclei at $\sqrt{s_{\rm NN}}$=193 GeV at RHIC have gathered a lot of attention recently. The STAR experiments at RHIC have reported interesting results on particle production as well as azimuthal flow of hadrons~\cite{star_data}. U+U collisions are of special interest  due to the non-spherical prolate shape of the colliding nuclei~\cite{uli_prl, uli_uu, nepali, nepali1, bjorn_uu, uli_new, janeda} and as a result, even the most central collisions can lead to different collision geometry and consequently  different values of  charged particle multiplicity and anisotropic flow parameters. 
Recently it has been reported that the most central events in U+U collisions can be identified from the spectator energy deposition at the Zero Degree Calorimeters (ZDCs). In addition, the multiplicity distribution of elliptic flow along with the ZDCs informations can be used to separate different orientations of U+U collisions~\cite{star_data}.

We know that photons are emitted throughout the life time of the evolving system and the thermal emission of photons is sensitive to the initial stages of the produced matter. Thus, photon production from different orientations of U+U collisions can provide valuable information about the hot and dense initial stage of the expanding system and also its evolution. In addition, it would be interesting to know how large is the photon $v_2$ originating from  fully overlapping U+U collision and if its comparison with the photon $v_2$ from non-central Au+Au collisions can help us to understand the photon $v_2$ puzzle.   

We calculate thermal photon spectra and differential elliptic flow at RHIC for two different orientations, {\it tip-tip} and {\it body-body} which are the limiting cases (of particle multiplicity) of fully overlapping U+U collisions.  In body-body  collisions  the major axes of the two incoming  uranium nuclei are perpendicular to the z axis (beam axis)  whereas for tip-tip collisions the major axes are parallel to the beam direction. The tip-tip collisions produce a circular overlapping zone on the transverse plane and the body-body collisions  lead to an elliptical shape and a larger size of the overlapping zone (see Fig 1). Although the number of participants in both these collisions are same, number of binary collisions is about 30\% larger for the tip-tip configuration. The energy density produced is larger and consequently a higher final charged  particle multiplicity is observed for tip-tip collisions than for the body-body collisions. However the  body-body collisions produce a large $v_2$ because of the initial geometry of the overlapping zone~\cite{star_data}. 

It has been shown in Ref.~\cite{uli_prl} that the value of  the initial spatial anisotropy ($\epsilon_{\rm in}$) for full overlap body-body collision is similar to the $\epsilon_{\rm in}$ calculated for Au+Au collisions at RHIC at an impact parameter $\sim$ 7 fm, however, the system produced in Au+Au collision  is about half of the size of system produced in U+U collisions. Thus, the photon spectra and elliptic flow from the different orientations of U+U collisions along with the Au+Au results at  RHIC would enrich our understanding of the hot and dense initial state produced in relativistic heavy ion collisions. We keep our calculations simple by using a hydrodynamical model with smooth initial density distribution. The initial energy depositions for the tip-tip and body-body orientations are taken from Ref.~\cite{uli_prl} and the  calculated thermal photon spectra and elliptic flow parameter depend strongly on the initial conditions. An event-by-event hydrodynamic model calculation including viscous effect is expected to provide a more accurate estimation of the photon spectra and elliptic flow parameter. However in the present study we are more interested in showing the qualitative difference in the spectra and $v_2$ resulting from the different orientations of the uranium nuclei in and also the potential of thermal photons from U+U collisions to be used as probe to study the relativistic heavy ion collisions. In addition, we calculate prompt photons from body-body and tip-tip collisions of uranium nuclei and compare the direct photon spectra (obtained by adding prompt and thermal contributions) for the two configurations.

In Section II we briefly discus the initial parameters and the framework for the model calculation. Thermal photon spectra and elliptic flow results are presented in Section III and in the next section we summarize the results. 
 
\begin{figure}
\centerline{\includegraphics*[width=8.0 cm]{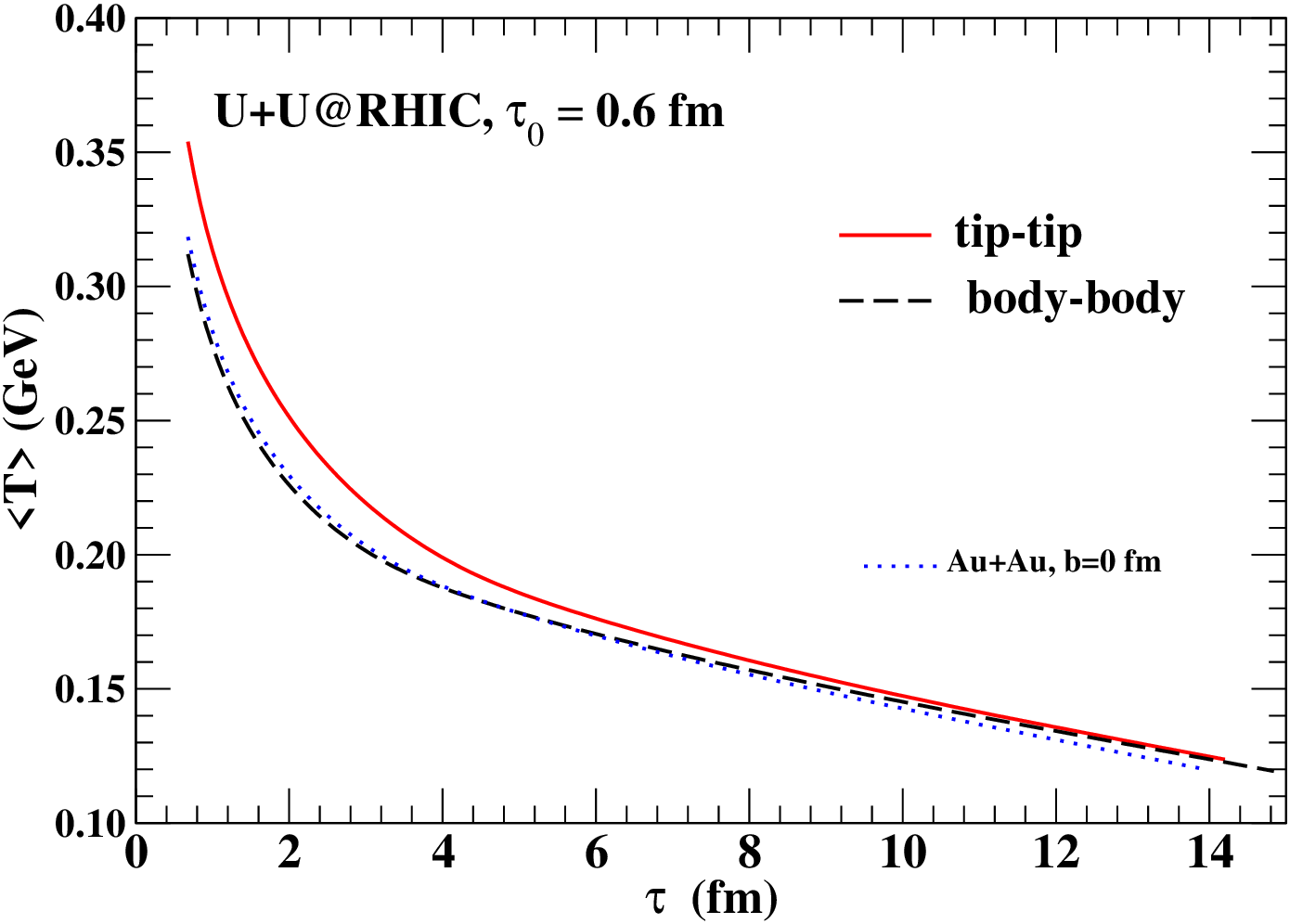}}
\centerline{\includegraphics*[width=8.0 cm]{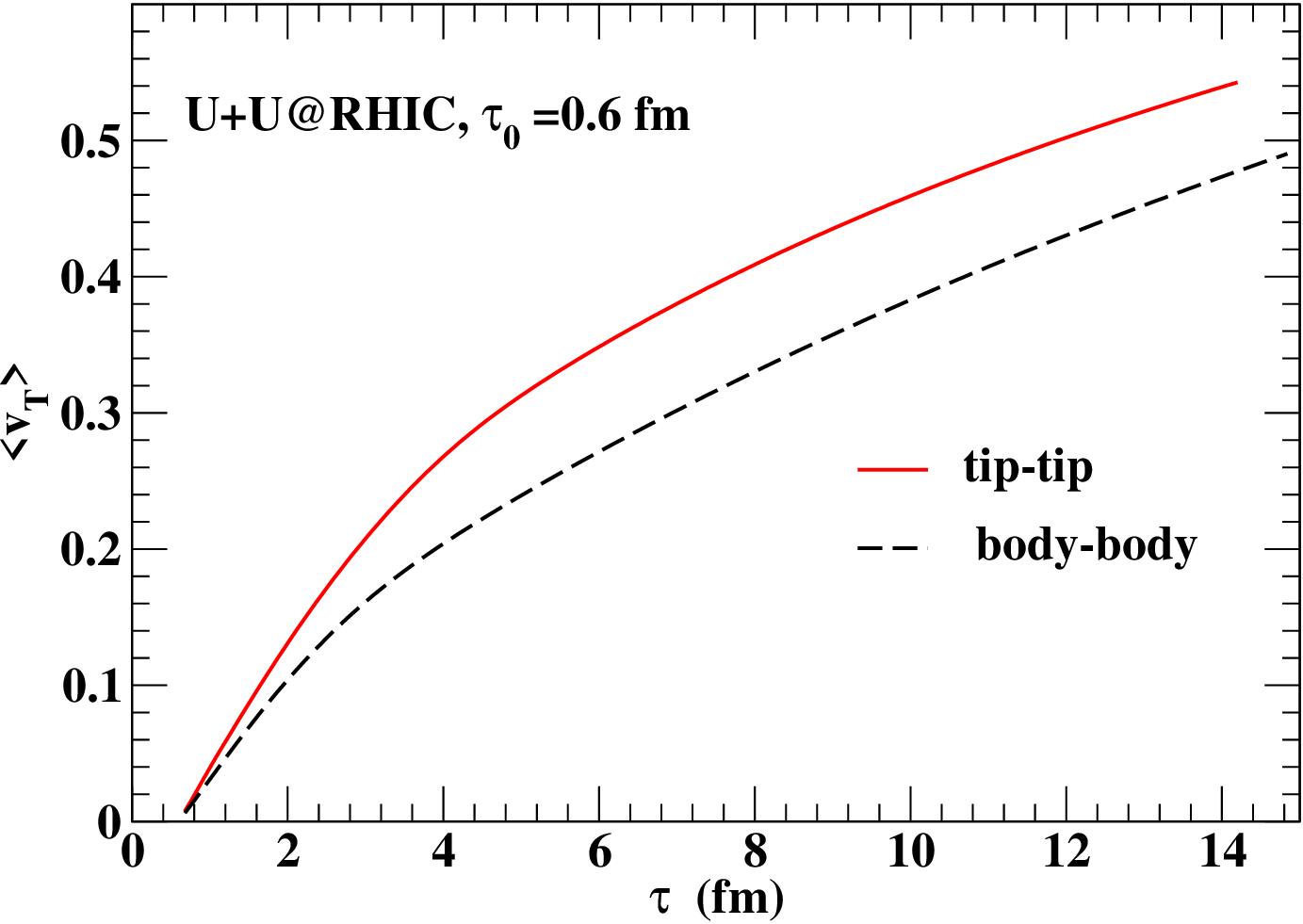}}
\caption{(Color online) Time evolution of (a) average temperature $\langle T \rangle$ and  (b) average transverse flow velocity $\langle v_T \rangle$ for tip-tip and body-body full overlap U+U collisions at RHIC. }
\label{vt}
\end{figure}

\section{Full overlap U+U collisions at RHIC}
We use Woods-Saxon parameterization for the nuclear  density distribution of deformed uranium nuclei of the form~\cite{uli_new}
\begin{equation}
 \rho (r, \theta)=\frac{\rho_0}{1+ {\rm exp} \frac {(r-R (\theta))} { \xi}} 
\end{equation}
where,
\begin{equation}
 R(\theta )=R_0 [1+ \beta_2 Y_{2}^{0}(\theta ) + \beta_4 Y_{4}^{0}(\theta )]
\end{equation}
The spherical harmonic functions and the $\beta$ values introduce the deformation from spherical shape in the uranium nucleus.  
Here $\beta_2$ and  $\beta_4$ are 0.28 and 0.093 respectively~\cite{uli_new}. $R_0$ is taken as 6.86 fm and $\xi$ is 0.44 fm~\cite{uli_new}. Using this parameterization in Optical Glauber Model we calculate the number of wounded nucleons $(N_{\rm part})$ and  binary collisions $(N_{\rm coll})$ for different orientations of full overlap  U+U collisions at RHIC. The value of $N_{\rm coll}$ is $\sim$ 1870 and $\sim$ 1430 for tip-tip and body-body collisions respectively, whereas  $N_{\rm part}$ is same for both the cases.

We modify the 2+1 dimensional longitudinally boost invariant hydrodynamic code AZHYDRO~\cite{uli} with smooth initial density distribution to study the evolution of the system produced in U+U collisions at RHIC. The initial formation time $\tau_0$ is considered as 0.6 fm. The corresponding initial entropy densities ($s_0$) at the center of the fireball are taken as 167 fm$^{-3}$ and 110 fm$^{-3}$ for full overlap tip-tip and body-body collisions respectively, and thus the value of $s_0$ is about 34\% higher for tip-tip configuration~\cite{uli_prl}.  For Au+Au collisions at 200A GeV, $s_0$ is taken as 117 fm$^{-3}$ and it reproduces the experimentally measured charged particle multiplicity at mid-rapidity.

A lattice based equation of state~\cite{eos} is used and the final freeze-out temperature $T_f$ is considered as 140 MeV. We check the sensitivity of our results to the initial parameters of the model calculation by changing the value of $\tau_0$ and $T_f$ from their default values. For initial density distribution we use both wounded nucleon profile ($\alpha=$0) as well as a two component ($\alpha=$0.25) model~\cite{uli_prl} (where  the initial entropy is taken as proportional to a linear combination of 25\%  of $N_{\rm coll}$ and 75\% of $N_{\rm part}$) to calculate the photon production from U+U collisions.

The nucleon-nucleon inelastic cross section $\sigma_{\rm NN}$  for 200 GeV collisions  is 42 mb and we use the same $\sigma_{\rm NN}$ for 193 GeV collisions of uranium nuclei at RHIC.  We assume that the small change in the value of  $\sigma_{NN}$ for change in centre of mass energy from 200 to 193 GeV would not affect our results significantly. We use  next-to-leading order QGP rates from~\cite{amy, nlo_thermal} to calculate the photons spectra and elliptic flow. The photon production from hadronic phase is calculated  using the parameterization given in ~\cite{trg} for different hadronic channels. The $p_T$ spectra are calculated by integrating the emission rates over the space-time 4-volume and the elliptic flow parameter $v_2$  is calculated using the relation:
\begin{equation}
v_2 (p_T) \ = \ \langle {\rm cos}(2\phi) \rangle \ = \ \frac{\int_0^{2\pi} \, d\phi \, {\rm cos}(2\phi) \, \frac{dN}{p_T dp_T dy d\phi}}{\int_0^{2\pi} \, d\phi  \, \frac{dN}{p_T dp_T dy d\phi}} \, .
\end{equation}
\begin{figure}
\centerline{\includegraphics*[width=8.0 cm]{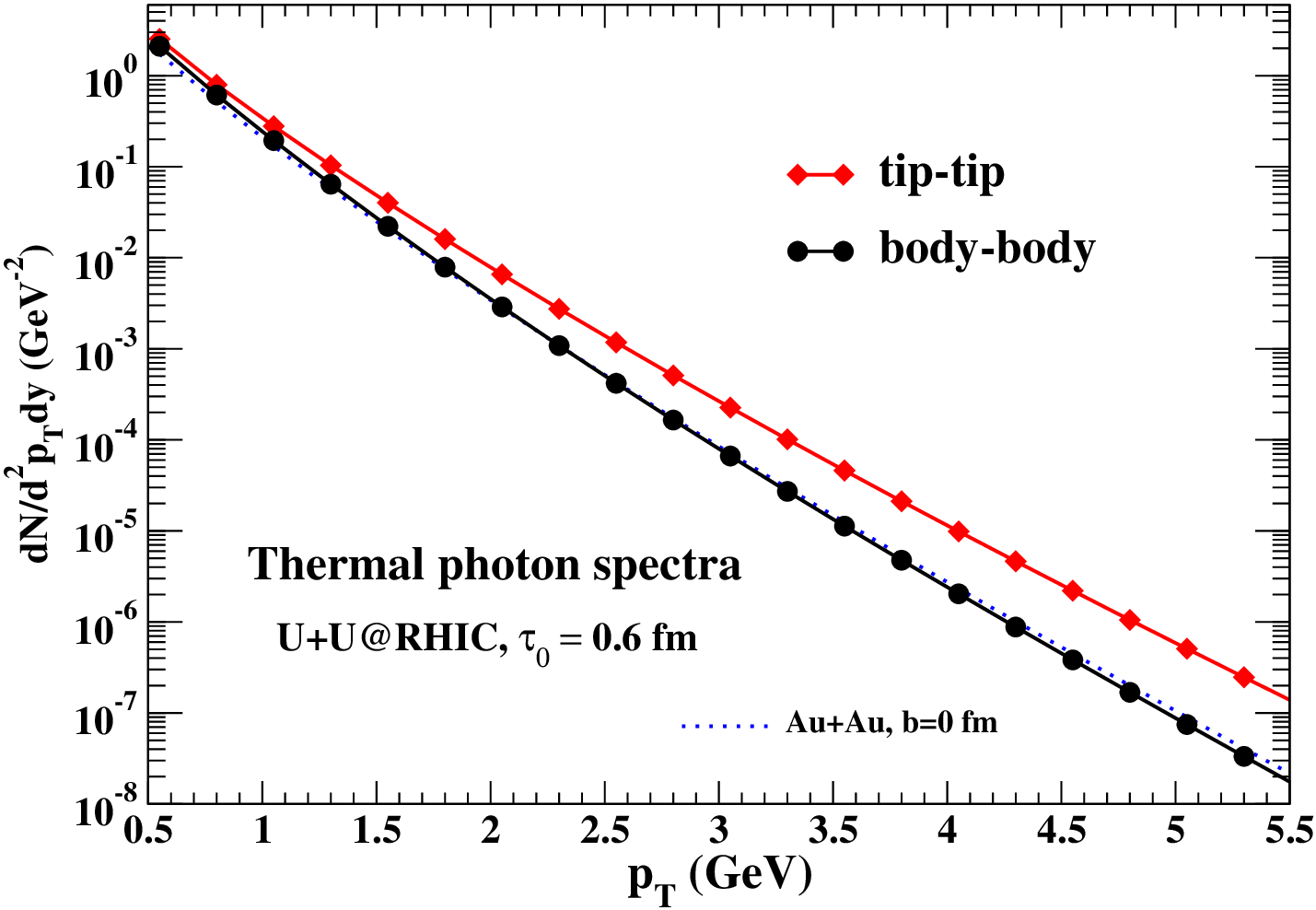}}
\centerline{\includegraphics*[width=8.0 cm]{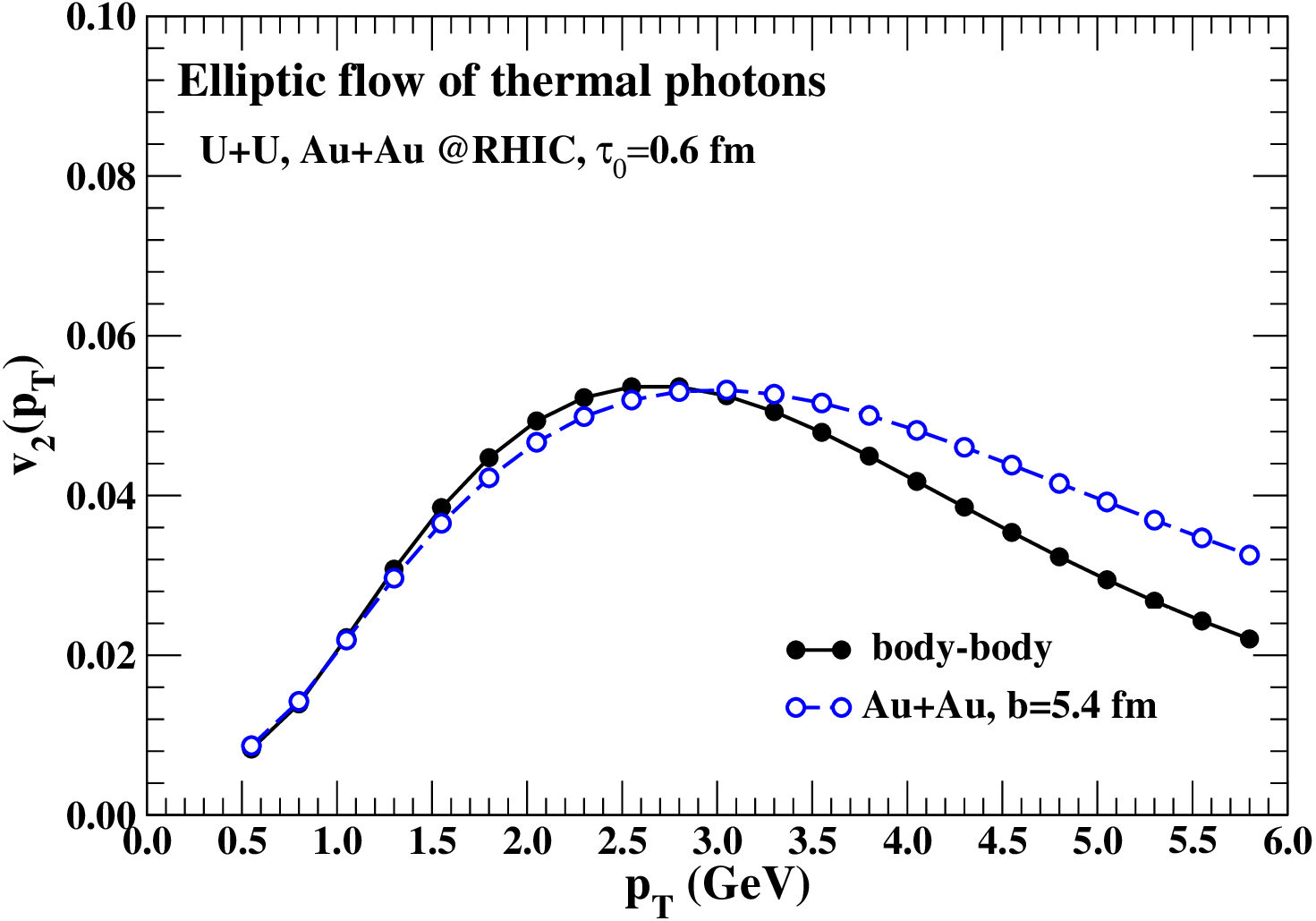}}
\caption{(Color online). Thermal photon (a) $p_T$ spectra and (b) elliptic flow from full overlap U+U collisions using hydrodynamic model for $\tau_0 =$ 0.6 fm and $\alpha=$0.25. }
\label{0.6}
\end{figure}

\section{Results}
The time evolution of  average temperature (upper panel) and average transverse flow velocity (lower panel)  for the two orientations of U+U collisions at RHIC are shown in Fig.~\ref{vt}. The averages are  obtained using the relation,
\begin{equation}
 \langle f \rangle = \frac{ \int \, dxdy \, \epsilon (x,y) f(x,y)}{ \int \, dxdy \, \epsilon (x,y)} \, .
\end{equation}
The value of $\langle T \rangle$ at time $\tau_0$ is $\sim$ 350 MeV for tip-tip collisions which is about 6\% larger than for body-body collisions. The $\langle T \rangle$ for most central Au+Au collisions is found to be close to that of body-body collisions as the initial entropy densities for these two cases are similar. 
 We also see that the  average $v_T$ is significantly larger for tip-tip collisions throughout the system evolution and the system life-time is slightly larger for body-body collisions.

The upper panel of Fig.~\ref{0.6} shows the thermal photon $p_T$ spectra  for full overlap tip-tip and body-body collisions of uranium nuclei considering initial formation time $\tau_0=$ 0.6 fm and $\alpha=$0.25. The $p_T$ spectrum from central Au+Au collisions is also shown in the figure for a comparison. Thermal photon production is found to depend strongly on the orientation of the colliding uranium nuclei.  The  production is significantly larger for tip-tip collisions in the higher $p_T$ ($>$ 1 GeV) region and photon spectrum from body-body orientation falls more rapidly compared to the tip-tip spectrum for larger values of $p_T$. One can see from the figure that the production for tip-tip collisions is about a factor of 2--5 times larger than body-body collisions in the region $2 < p_T < 4$ GeV.  We have discussed that the produced fireball in tip-tip collision is smaller in size and has larger initial energy and/or entropy density and temperature than the body-body configuration.  Higher initial temperature results in more high $p_T$ photons  from the initial stages in tip-tip collision which make the spectrum flatter. The production in the low $p_T$ ( $< $ 1 GeV) region for body-body as well as for tip-tip collisions is mostly from the hadronic phase. Any other orientation of full overlap U+U collision would result in photon spectra lying in between the spectra from tip-tip (upper limit) and body-body (lower limit) collisions in the high $p_T$ region.

 It is to be noted that the results presented here depend strongly on the initial energy deposition values taken from Ref.~\cite{uli_prl} for the two limiting configurations of the uranium nuclei. A more realistic estimation of the photon spectra and elliptic flow parameter demands these initial conditions to simultaneously reproduce the experimental charged particle spectra and anisotropic flow parameter. However, this seems little difficult at the moment due to the present status of the available experimental data.  In this study we mainly focus on thermal photons as a potential probe to study U+U collisions at RHIC and the qualitative nature of the results presented here is expected to remain unchanged for small changes in the value of initial energy deposition.

\begin{figure}
\centerline{\includegraphics*[width=8.0 cm]{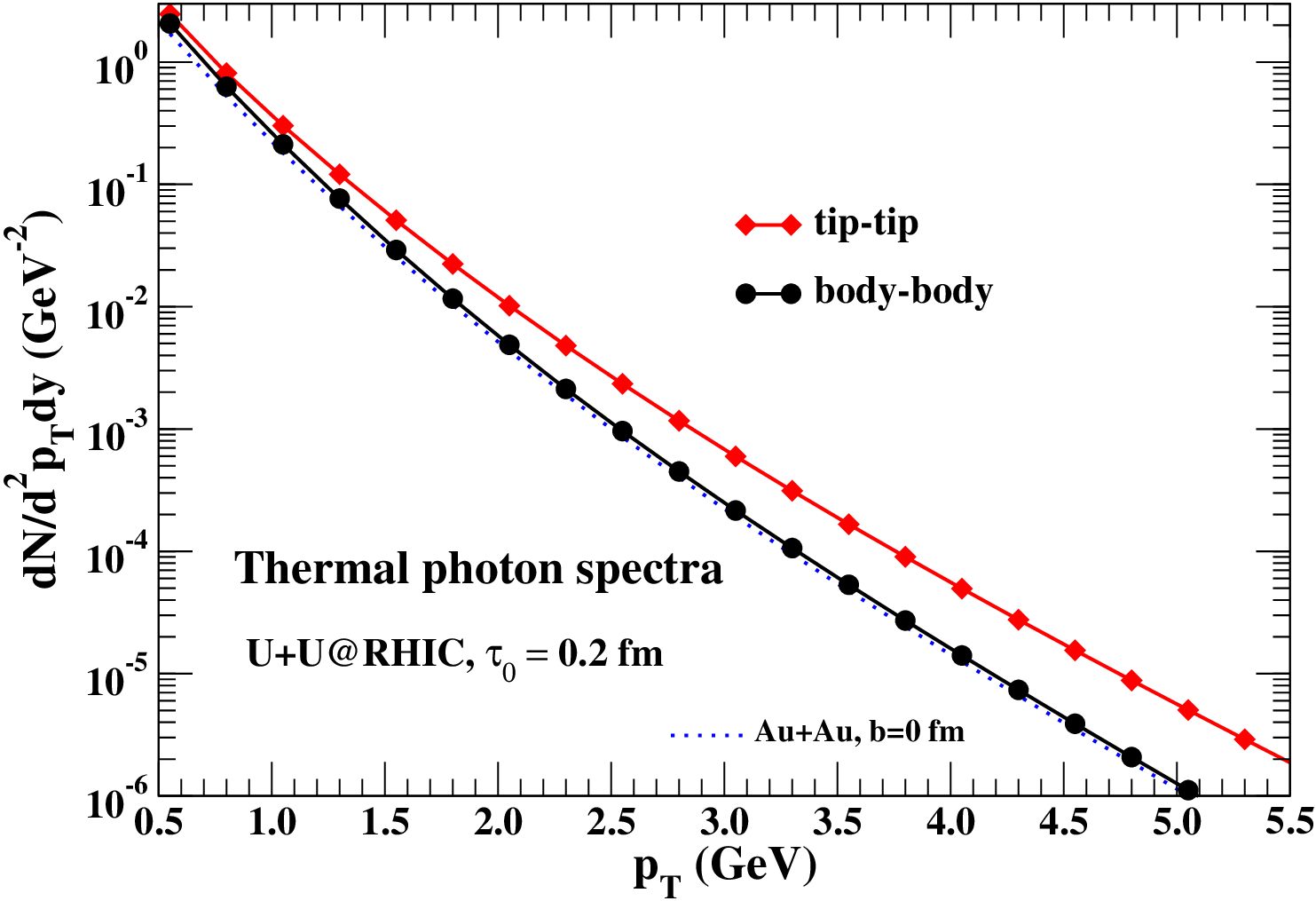}}
\centerline{\includegraphics*[width=8.0 cm]{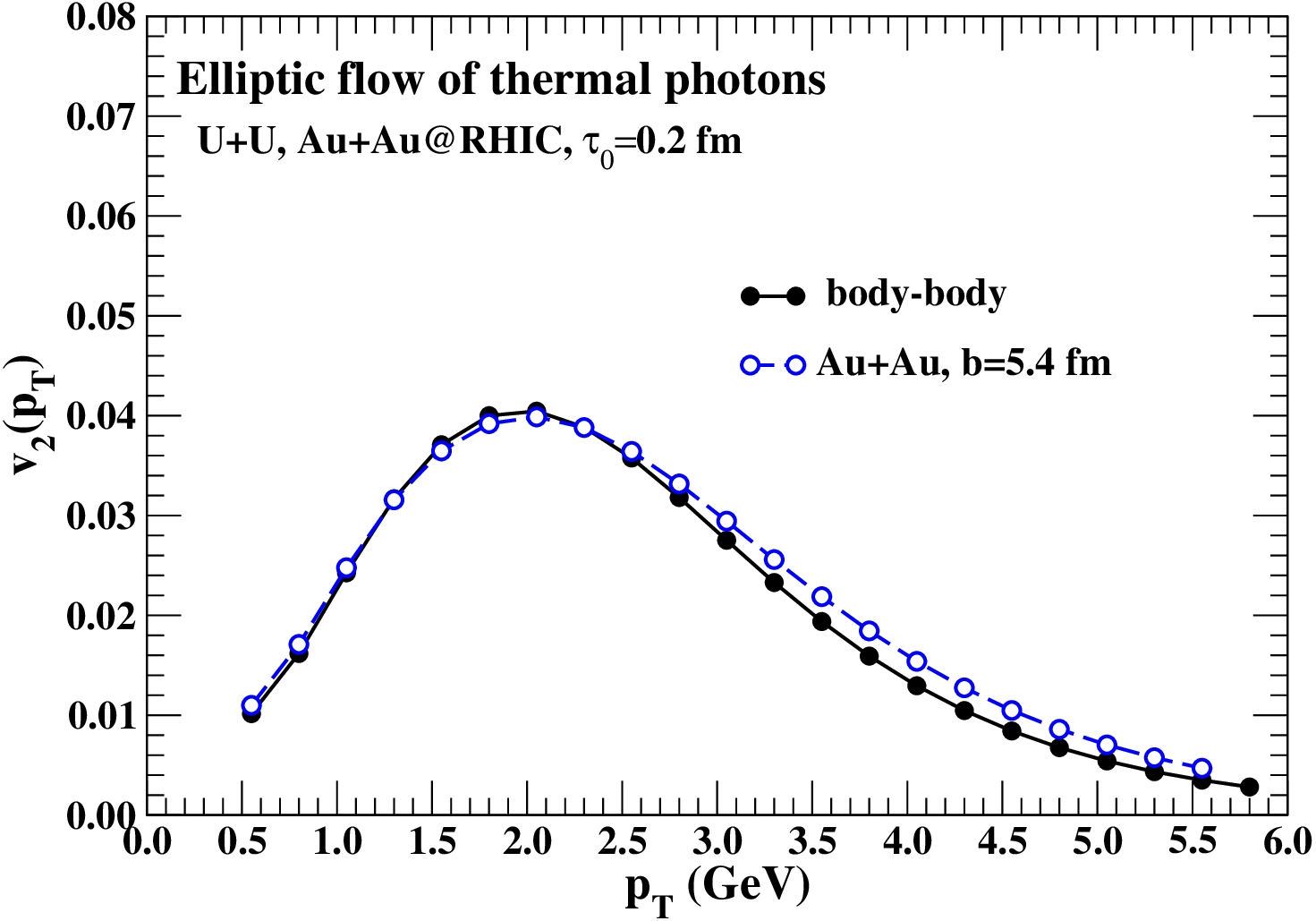}}
\caption{(Color online) Thermal photon (a) $p_T$ spectra and (b) elliptic flow from full overlap U+U collisions using hydrodynamic model for $\tau_0 =$ 0.2 fm and $\alpha=$0.25. }
\label{spec_0.2}
\end{figure}

The elliptic flow parameter $v_2(p_T)$ for body-body collisions is shown in lower panel of Fig.~\ref{0.6}. The $v_2(p_T)$ for tip-tip collisions is zero as there is no initial spatial anisotropy present in the system (It is to be noted that hydrodynamical model calculation using fluctuating initial conditions would result in very small but non-zero photon elliptic flow even for tip-tip collisions of uranium nuclei).
However, we see significantly large elliptic flow for body-body collisions. In addition, this large flow result is found to be close to the $v_2(p_T)$ calculated from Au+Au collisions at RHIC at an impact parameter b=5.4 fm. The initial spatial anisotropy  of the overlapping zone is calculated using the relation,  
\begin{equation} 
\epsilon_{\rm {in}} = \frac { \int dx dy  \ \epsilon (x,y, \tau_0) (y^2-x^2) }{ \int dx dy \ \epsilon (x, y, \tau_0) (y^2+x^2)} \,
\end{equation}
where, $\epsilon (x, y, \tau_0)$ is the energy density at point (x,y) on the transverse plane at time $\tau_0$. It is to be noted that the initial spatial  anisotropy of the overlapping zone for full overlap body-body collision is  about 0.26, whereas the value of $\epsilon_{\rm in}$ is about 0.19 at b=5.4 fm for Au+Au collisions. The peak of $v_2(p_T)$  appears around $p_T\sim$ 2 GeV and the competing contributions of photons originating  from the different stages of the evolving system determine the shape of the  $v_2(p_T)$ curve. As the relative contribution from the hadronic phase compared to QGP phase for mid-central Au+Au collisions is much larger than for body-body collisions of uranium nuclei, we see the results in lower panel of Fig.~\ref{0.6} are similar even for a smaller $\epsilon_{\rm in}$ in case of Au+Au collisions.

We know that  photon $v_2(p_T)$ rises towards peripheral collisions as the initial spatial anisotropy increases (as in the case for the elliptic flow of hadrons) and also due to change in the relative contributions from the quark matter and  hadronic matter phases~\cite{cfhs}. The body-body collision of uranium nuclei  shows  large elliptic flow  even for most central collisions and thus it would be interesting to see if $v_2$ for this orientation increases significantly towards peripheral collisions.

\begin{figure}
\centerline{\includegraphics*[width=8.0 cm]{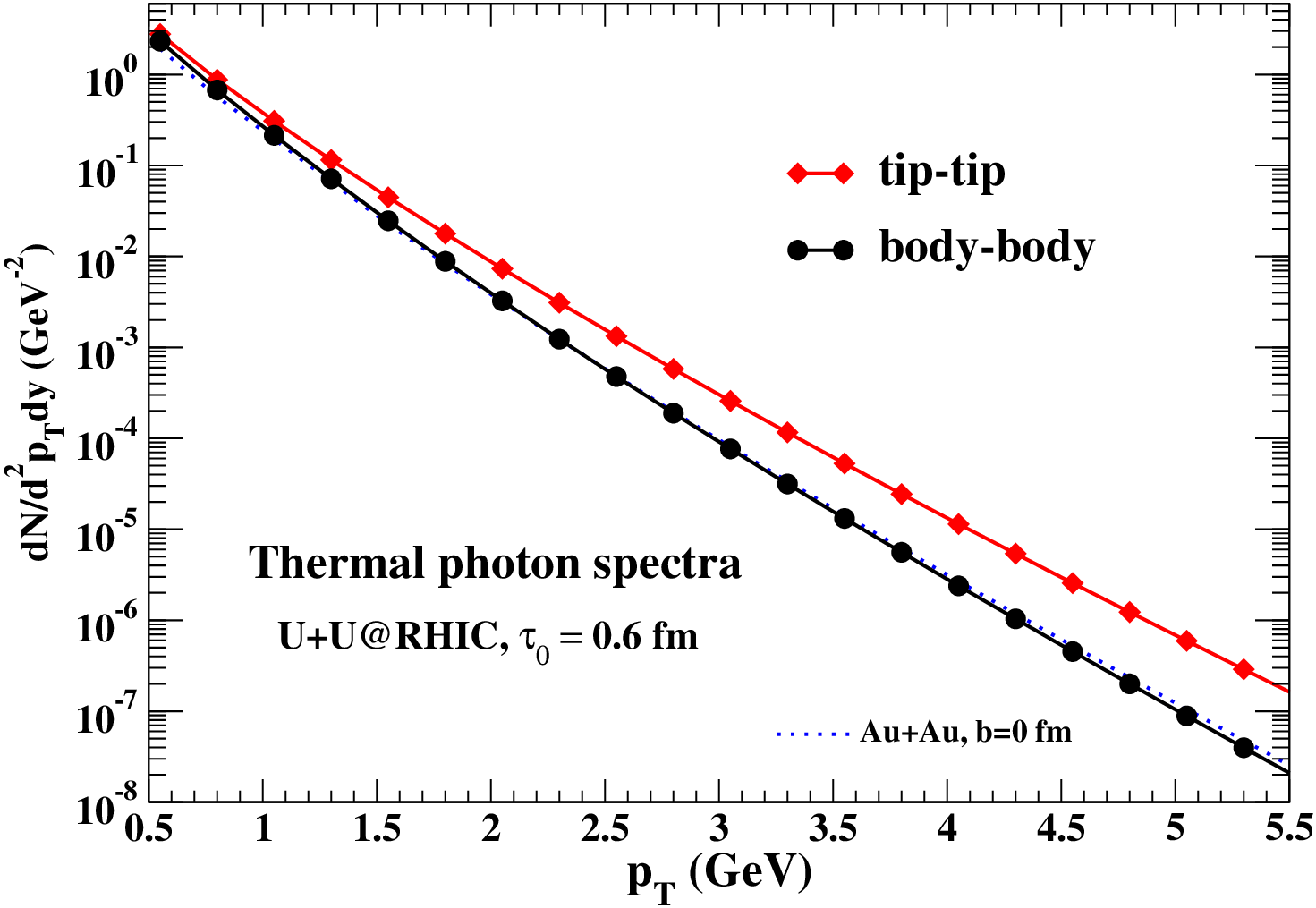}}
\centerline{\includegraphics*[width=8.0 cm]{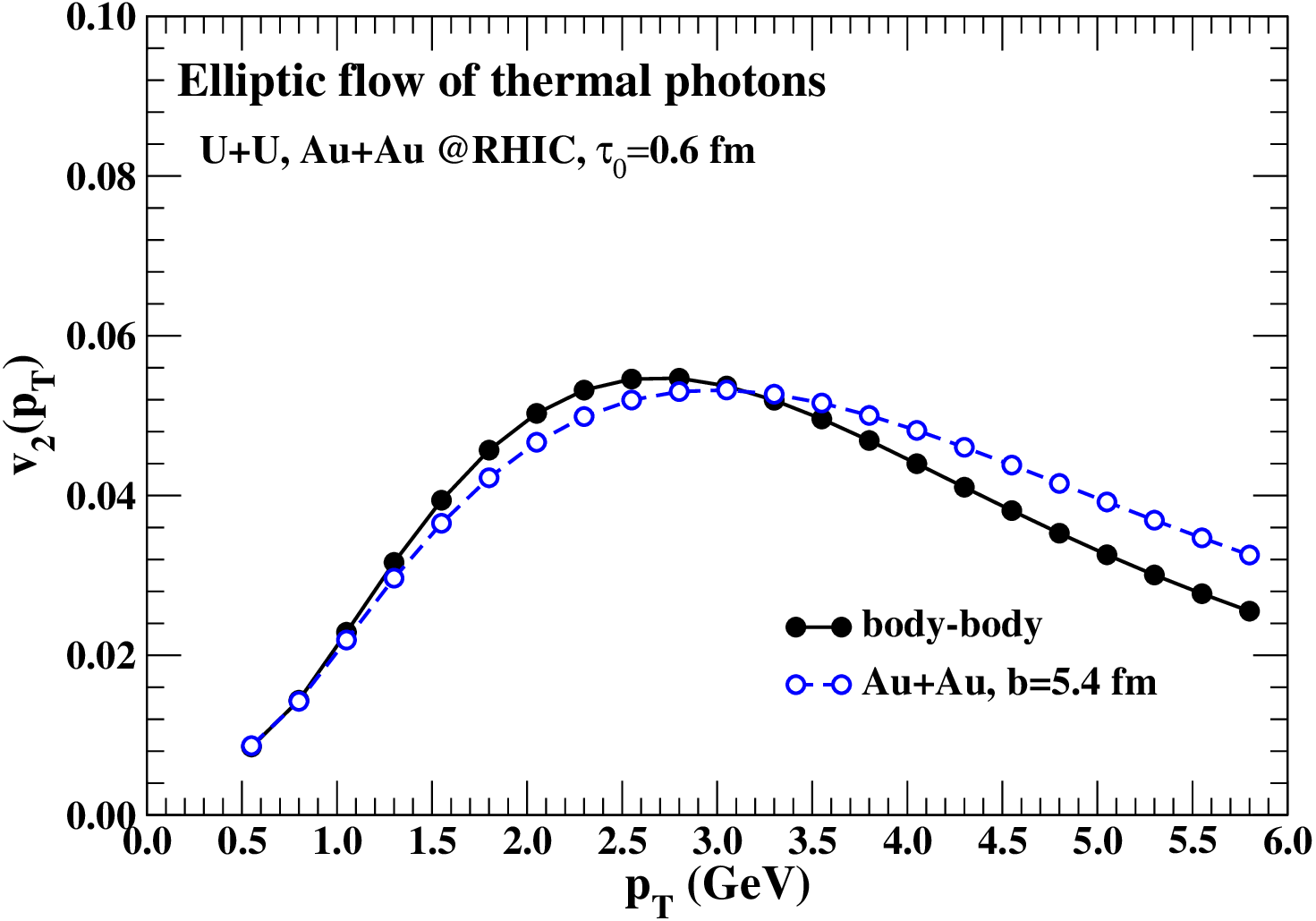}}
\caption{(Color online). Thermal photon (a) $p_T$ spectra and (b) elliptic flow from full overlap U+U collisions using hydrodynamic model for $\tau_0 =$ 0.6 fm and $\alpha=$0. }
\label{0.6_alpha_0}
\end{figure}

We recall that the initial formation time $\tau_0$ plays important role in photon calculations as a smaller value of $\tau_0$ means larger initial temperature and more production of high $p_T$ photons~\cite{cs,chre1}.  Thermal photon spectra and $v_2$ for $\tau_0$ $=$ 0.2 fm are shown in Fig 4. The value of $\tau_0$ is reduced from 0.6 to 0.2 fm, keeping the total entropy of the system fixed. We see  enhanced production of thermal photons compared to $\tau_0=$ 0.6 fm both for tip-tip and body-body collisions (upper panel of Fig. 4). However, the difference between the slopes of the spectra for the two orientations remain similar to the results obtained at  $\tau_0$ = 0.6 fm. 
Photon $v_2$ for full overlap tip-tip collisions is zero and does not depend on the initial parameters of the hydrodynamic calculation. However, for body-body collisions we see large elliptic flow (lower panel of Fig. 4) and again the result is close to the photon $v_2$ calculated from  Au+Au collisions at RHIC at b=5.4 fm and at $\tau_0$=0.2 fm. The thermal photon spectra and elliptic flow for $\tau_0 =$ 0.6 fm and $\alpha=$0  are shown in Fig~\ref{0.6_alpha_0}. The elliptic flow results from U+U as well as from the Au+Au collisions are found to be somewhat larger compared to the results obtained by considering $\alpha=$0.25. However, the qualitative nature of the spectra as well as $v_2$ do not show strong dependence on the value of $\alpha$. We have also checked that the qualitative nature of the spectra and elliptic flow results presented here do not change significantly when the freeze-out temperature is reduced from 140 to 120 MeV. 

\begin{figure}
\centerline{\includegraphics*[width=8.0 cm]{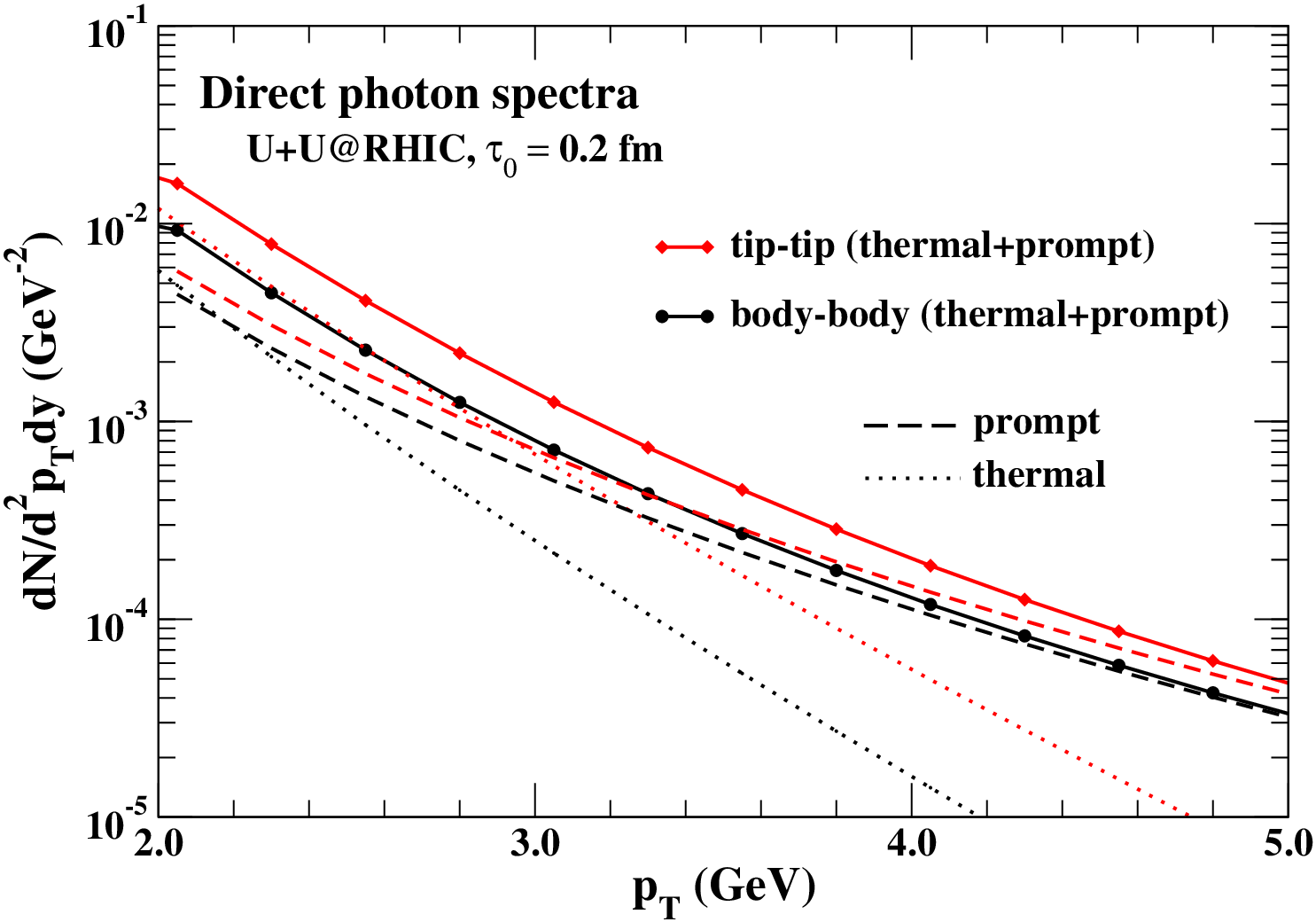}}
\caption{(Color online) Direct photon (thermal and prompt) spectra from full overlap U+U collisions.}
\label{prompt_spec}
\end{figure}

We know that the prompt photons produced in initial hard scatterings start to dominate the direct photon spectrum in the region  $p_T >$ 3 -- 4 GeV. We estimate the prompt photons~\cite{aurenche} using NLO pQCD calculation and CTEQ5M~\cite{cteq5m} structure function for the two limiting cases discussed here for full overlap U+U collisions at 193A GeV. As the value of $N_{\rm coll}$ is about 30\% larger for tip-tip than for the body-body configuration, the  prompt contribution is also found to be about 30\% larger for the tip-tip case (see Fig.~\ref{prompt_spec}). One can see from the figure that the direct photon spectrum (combining prompt and thermal contributions) for tip-tip configurations is about a factor of 2 larger than for the body-body collisions in the range $p_T < $ 5 GeV. Thus, we see that the direct photon spectra from full overlap U+U collisions at RHIC show significant  dependence on the orientation of the colliding nuclei even at larger values of $p_T$ ($\sim$4 -- 5 GeV) where the non-thermal contributions dominates the spectra.

It is to be noted that fluctuations in the initial density distribution might result in a small increase in $v_2$ in the high $p_T$ region for body-body collisions and also a small but non-zero $v_2$ even for tip-tip collisions~\cite{chre}. In addition,  viscosity plays a role in photon $v_2$ calculations by reducing the $v_2$ at higher $p_T$~\cite{phot_v2_shen}. Thus, a complete calculation using viscous hydrodynamics with event-by-event fluctuating initial condition would be valuable and we postpone this for a future study~\cite{pingal1}. However, the results presented in this paper are believed to be generic in nature and should remain unaltered even with the modifications discussed above.

We know that the different orientations of the most central U+U collisions can be distinguished from the spectator energy deposition at the ZDCs together with the particle multiplicities. Thus, experimental determination of photon $v_2$ from different orientations of uranium nuclei should also be possible. 

We see a significant enhancement in the photon production from tip-tip  U+U collisions compared to central Au+Au collisions. In addition, the  photon $v_2(p_T)$ from the body-body U+U collisions is found to be similar to the elliptic flow from mid central Au+Au collisions at RHIC using hydrodynamical model calculation. However, it is to be noted that the system produced in mid-central Au+Au collisions and in body-body U+U collisions are very different in terms of initial temperature, system size and life-time. It is shown that the time evolution of average temperature for central Au+Au collisions and body-body U+U collisions are close to each other. Thus, the system produced in Au+Au collisions at b=5.4 fm is expected to have smaller temperature than the one  in body-body U+U collisions.  As a result, the relative contributions of the QGP and hadronic matter phases to the total photon $v_2$ are very different although the flow results look similar in those two cases. Now, it is not possible to know the separate contributions of the QGP and  the hadronic phases to photon elliptic flow from the experimentally obtained $v_2$ data. However, theory calculation has this advantage which helps us to understand that two very different system (with different relative contributions from quark and hadronic matter phases) can have similar $v_2$.

Thus, experimental determination of photon spectra and elliptic flow from U+U collisions at RHIC would be valuable and comparison of the results with the photon results from Au+Au collisions at various centrality bins would provide an additional handle to study  photon production in relativistic heavy ion collisions.

\section{summary}
We have calculated $p_T$ spectra and differential elliptic flow $v_2(p_T)$ of thermal photons for tip-tip and body-body orientations of full overlap U+U collisions at RHIC using hydrodynamic model with smooth initial density distribution. We see significantly larger production of thermal photons from tip-tip collisions in the region $p_T >$ 1 GeV compared to the body-body collisions. The results depend on the difference in energy depositions (the values of which  are yet to be constrained precisely from hadronic measurements) for the two limiting configurations of uranium nuclei. Larger initial energy densities as well as temperatures for tip-tip collisions result in more high $p_T$ photons from the early stage of system evolution. 
We see relatively larger production of prompt photons from the tip-tip collisions than from the body-body collisions (as $N_{\rm coll}$ is larger for tip-tip collision) and  thus, the direct photon spectra obtained by adding the prompt and thermal contributions also show significant difference between the limiting cases of full overlap U+U collisions upto a large $p_T$ ($\sim$ 5 GeV).
Photon $v_2$ from tip-tip collisions is close to zero from hydrodynamic calculation as there is no spatial anisotropy present in the system (it is to be noted that fluctuations in the initial density distribution would result in small $v_2$ even for tip-tip collisions.) On the other hand, we see significantly large photon $v_2$ from full overlap body-body collisions which is comparable to the photon $v_2$ calculated at b=5.4 fm from 200A GeV Au+Au collisions at RHIC.  Comparison of photon $v_2$ from body-body U+U collisions and from mid central Au+Au collisions at RHIC would be valuable to understand the photon $v_2$ puzzle. We also calculate the spectra and elliptic flow parameter from U+U and Au+Au collisions by changing the initial parameters of the hydrodynamic model calculation and see that the qualitative nature of the results remain unchanged.

\begin{acknowledgments} 
One of us (DKS) gratefully acknowledges the grant of Raja Ramanna Fellowship by the Department of Atomic Energy, India. We thank Dr. Somnath De for useful discussions and for the prompt photon calculation.
\end{acknowledgments}


\begin{thebibliography}{99}
\bibitem{uli} P. Kolb and U Heinz, 
 Hydrodynamic description of ultrarelativistic heavy ion collisions
Quark gluon plasma 3, R. C. Hwa (ed.) {\it et al.}, p634 (2003).

\bibitem{hydro1}P. Huovinen, in Quark Gluon Plasma 3 , edited by R. C. Hwa and X. N. Wang (World Scientific, Singapore,  2004),  p. 600 [nucl-th/0305064];  P. F. Kolb, J. Sollfrank, and U. Heinz Phys. Rev. C {\bf 62}, 054909 (2000);  D. Teaney, J. Lauret, E.V. Shuryak, arXiv:0110037 [nucl-th];  P.~Huovinen and P.~V.~Ruuskanen, Ann.\ Rev.\ Nucl.\ Part.\ Sci.\  {\bf 56}, 163 (2006);  P.~Romatschke and U.~Romatschke, Phys.\ Rev.\ Lett.\  {\bf 99}, 172301 (2007);  D.~A.~Teaney,  arXiv:0905.2433 [nucl-th].



\bibitem{hannu} H. Holopainen, H. Niemi, and K. J. Eskola, Phys. \ Rev. \ C {\bf 83}, 034901 (2011). 

\bibitem{pt}B. Schenke, P. Tribedy, and  R. Venugopalan,
Phys. Rev. Lett. {\bf 108}, 252301 (2012).

\bibitem{scott}  U. Heinz, Z. Qiu, and C. Shen, Phys.\ Rev. C {\bf 87}, 034913 (2013).

\bibitem{hannah} C. E. Coleman-Smith, H. Petersen, and R. L. Wolpert, J. \ Phys. \ G {\bf 40}, 095103  (2013).

\bibitem{sorenson} P. Sorensen, J. \ Phys. \ G {\bf 37}, 094011 (2010).

\bibitem{nex} J. Takahashi {\it et al}, Phys.  \ Rev. \ Lett. {\bf 103} 242301, (2009).


\bibitem{alver} B. Alver and G. Roland, \ Phys. \ Rev. \ C {\bf 81}, 054905 (2010).

\bibitem{flow_phenix} A. Adare  {\it et al.} [PHENIX Collaboration] Phys. \ Rev. \ Lett. {\bf 107}, 252301 (2011). 

\bibitem{flow_lhc} Z. Qiu, C. Shen, and  U. Heinz, \ Phys. \ Lett. {\bf B707}, 151 (2012).

\bibitem{flow_atlas} G. Aad {\it et al.} [ATLAS Collaboration], \  Phys. \ Rev. C {\bf 86} 014907 (2012).



\bibitem{phot} 
  P.~V.~Ruuskanen,
  Nucl.\ Phys.\  A {\bf 544}, 169 (1992), and references therein; 
 D. K. Srivastava, \ J. \ Phys. \ G {\bf 35}, 104026 (2008).

\bibitem{phenix_phot} A. Adare {\it et al. } [PHENIX Collaboration]  Phys. \ Rev. \ Lett. \ {\bf 104}, 132301 (2010), A. Adare {\it et al.} [PHENIX Collaboration],  Phys. \ Rev. \ C {\bf 91}, 064904 (2015).  

\bibitem{alice_phot} M. Wilde for the ALICE Collaboration, Nucl. Phys. {\bf A 904-905}, 573c (2013); J.~Adam {\it et al.} [ALICE Collaboration],
  Phys.\ Lett.\ B {\bf 754}, 235 (2016).


\bibitem{cfhs} R. Chatterjee, E. S. Frodermann, U. Heinz, and D. K. Srivastava, Phys. \ Rev. \ Lett. {\bf 96}, 202302  (2006).


\bibitem{phenix_v2} A. Adare {\it et al. } [PHENIX Collaboration]  Phys. \ Rev. \ Lett. \ {\bf 109}, 122302 (2012);  A. Adare {\it et al.} [PHENIX Collaboration] Phys. \ Rev. \ C {\bf 94}, 064901 (2016).


\bibitem{alice_v2} D. Lohner for the ALICE Collaboration, J. \ Phys. \ Conf. \ Ser. {\bf 446}, 012028 (2013).




\bibitem{chre2} R. Chatterjee, H. Holopainen, I. Helenius, T. Renk, K. J. Eskola,  Phys.\ Rev.\  {\bf C88}, 034901 (2013).



\bibitem{cs} R. Chatterjee and D. K. Srivastava, Phys. \ Rev. \ C {\bf 79}, 021901(R) (2009); R. Chatterjee and D. K. Srivastava, Nucl. \ Phys. {\bf A830}, 503c (2009). 


\bibitem{hannu_phot} H. Holopainen, S. S.  Rasanen, and  K. J. Eskola,	
Phys. \ Rev. \ C {\bf84}, 064903 (2011).


\bibitem{phot_v2_hees} H. V. Hees, M. He,  and R. Rapp,  Nucl. \ Phys. \  {\bf A933}, 256 (2015).

\bibitem{dusling} K.~Dusling,  Nucl.\ Phys.\ A {\bf 839}, 70 (2010).


\bibitem{maxim} M. Dion, J.-F. Paquet, B. Schenke, C. Young, S. Jeon, and C. Gale,Phys. Rev. C {\bf 84}, 064901 (2011).

\bibitem{phot_v2_shen} C.~Shen, U.~W.~Heinz, J.~F.~Paquet and C.~Gale, Phys.\ Rev.\ C {\bf 89}, 044910 (2014); C. Shen, U. Heinz, J.-F. Paquet, I. Kozlov, and C. Gale, Phys. \ Rev. \ C {\bf 91}, 024908 (2015).

\bibitem{v22} H. van Hees, C. Gale, and R. Rapp,
Phys. Rev. C {\bf 84}, 054906 (2011).

\bibitem{v23} O. Linnyk, W. Cassing, and E. L. Bratkovskaya,
Phys. Rev. C {\bf 89}, 034908 (2014).

\bibitem{v24} F.-M. Liu and S.-X. Liu, Phys. Rev. C {\bf 89}, 034906 (2014).

\bibitem{v25} C. Gale, Y. Hidaka, S Jeon, S. Lin, J.-F. Paquet, R. D. Pisarski,
D. Satow, V. V. Skokov, and G. Vujanovic, Phys. Rev. Lett. {\bf 114}, 072301 (2015).

\bibitem{v26} B. Muller, S.-Y. Wu, and D.-L. Yang, 
Phys. Rev. D {\bf 89}, 026013 (2014).

\bibitem{v27} A. Monnai,  Phys. Rev. C {\bf 90}, 021901 (2014).

\bibitem{v28} L. McLerran and B. Schenke,  Nucl. Phys. A {\bf 929}, 71 (2014); L.~McLerran and B.~Schenke,
  Nucl.\ Phys.\ A {\bf 946}, 158 (2016).


\bibitem{v29} M. Greif, F. Senzel, H. Kremer, K. Zhou, C. Greiner, and Z. Xu,
 \ Phys. \  Rev. \ C {\bf 95}, 054903 (2017).

\bibitem{v30} I. Iatrakis, E. Kiritsis, C. Shen, D.-L. Yang,  J. \ High \ Energ. \ Phys. {\bf 04}, 035 (2017).

\bibitem{v31} V. Vovchenko, Iu. A. Karpenko, M. I. Gorenstein, L. M. Satarov, I. N. Mishustin, B. Kampfer, and H. Stoecker, \ Phys. \ Rev. \ C {\bf 94}, 024906 (2016).

\bibitem{v32} J.-F. Paquet, C. Shen, G. S. Denicol, M. Luzum, B. Schenke, S. Jeon, and C. Gale, \ Phys. \ Rev. \ C {\bf 93}, 044906 (2016).


\bibitem{star_data} L. Adamczyk {\it et. al.} [STAR Collaboration], \ Phys. \ Rev. \ Lett. {\bf 115}, 222301 (2015).

\bibitem{uli_prl} U. Heinz and A. Kuhlman, \ Phys. \ Rev. \ Lett. {\bf 94}, 132301 (2005).

\bibitem{uli_uu} A. Kuhlman and U. Heinz, Phys. \ Rev. \ C {\bf 72}, 037901 (2005).

\bibitem{nepali} C. Nepali, G. Fai and D. Keane, Phys. \ Rev. \ C {\bf 73}, 034911 (2006).

\bibitem{nepali1}  C.  Nepali,  G.  I.  Fai,  and  D.  Keane,  Phys. \ Rev. \  C {\bf 76}, 051902 (2007);  [Erratum:  Phys.  \ Rev.  \ C {\bf 76} , 069903 (2007)].

\bibitem{bjorn_uu} B. Schenke, P. Tribedy, and R. Venugopalan, \ Phys. \ Rev.\ C {\bf 89}, 064908 (2014).

\bibitem{uli_new} A. Goldschmidt, Z.  Qiu, C. Shen, and  U.  Heinz, \ Phys. \ Rev.\ C {\bf 92}, 044903 (2015).

\bibitem{janeda} S. Chatterjee, S. K. Singh, S. Ghosh, M. Hasanujjaman, J. Alam, and S. Sarkar, \ Phys. \ Lett. \ B {\bf 758}, 269 (2016).


\bibitem{eos}
M.~Laine and Y.~Schroder,
\newblock Phys. Rev. {\bf D73}, 085009 (2006).


\bibitem{nlo_thermal} J. Ghiglieri, J. Hong, A. Kurkela, E. Lu, G. D. Moore, and
D. Teaney, \ JHEP {\bf 1305}, 010 (2013).

\bibitem{amy} P.~Arnold, G.~D.~Moore, and L.~G.~Yaffe, JHEP {\bf 0112}, 009
(2001).


\bibitem{trg} S.~Turbide, R.~Rapp, and C.~Gale, \ Phys. \ Rev. \ C {\bf 69},
 014903 (2004).

\bibitem{chre1} R. Chatterjee, H. Holopainen, T. Renk, and K. J. Eskola,
Phys. \ Rev. C {\bf 85},  064910 (2012).


\bibitem{aurenche} P. Aurenche, M. Fontannaz, J.-P. Guillet, B. A. Knielhl, E. Pilon, and M. Werlen, \ Eur.  \ Phys. \ Jour. \ C {\bf9}, 107 (1999).

\bibitem{cteq5m} H. L. Lai, J. Huston, S. Kuhlmann, J. Morfin, F. Olness, J. F. Owens, J. Pumplin, W. K. Tung, \ Eur. \ Phys. \ J.  {\bf C12}, 375 (2000).

\bibitem{chre} R.~Chatterjee, H.~Holopainen, T.~Renk, and K.~J.~Eskola,  Phys.\ Rev. \ C {\bf 83}, 054908 (2011);  R.~Chatterjee, H.~Holopainen, T.~Renk, K.~J.~Eskola, J. Phys. G. Nucl. Part. Phys. {\bf 38}, 124136 (2011).



\bibitem{pingal1} P. Dasgupta, R. Chatterjee, and D. K. Srivastava [in preparation].







\end{thebibliography}
\end{document}